\documentclass[prd,,aps,nofootinbib,amsmath,amssymb]{revtex4}
\usepackage{amssymb,amsmath,epsfig,bm}
\usepackage[usenames,dvipsnames]{color}
\usepackage{mathrsfs}
\usepackage[bookmarks]{hyperref}

\newcommand{\Scri}{\mathscr{I}}
\newcommand{\sinn}{\mbox{sinn}}

\begin{document}

\title{The gravitational redshift of photons traversing a collapsing dust cloud and observable consequences}

\author{N\'estor Ortiz and Olivier Sarbach}
\affiliation{Perimeter Institute for Theoretical Physics, 31 Caroline St., Waterloo, ON, N2L 2Y5, Canada. \\Instituto de F\'isica y Matem\'aticas, Universidad Michoacana de San Nicol\'as de Hidalgo, Edificio C-3, Ciudad Universitaria, 58040, Morelia, Michoac\'an, M\'exico.}
\email{nortiz@perimeterinstitute.ca and sarbach@ifm.umich.mx}

\begin{abstract}
We analyze the frequency shift of photons propagating on an asymptotically flat spacetime describing a collapsing, spherical dust cloud. We focus on the case where the interaction of the photons with the matter can be neglected. Under fairly general assumptions on the initial data characterizing the collapse, we show that photons with zero angular momentum which travel from past to future null infinity, crossing the collapsing cloud through its center, are always redshifted with respect to stationary observers. We compute this redshift as a function of proper time of a distant stationary observer and discuss its dependency on the mass distribution of the cloud. Possible implications of this redshift effect for weak cosmic censorship and light propagation in cosmological spacetimes are also briefly discussed.
\end{abstract}

\maketitle

\section{Introduction}

The study of electromagnetic radiation emitted from the surface of a collapsing object is an important topic. Much information about the object can be obtained by the observable characteristics of this radiation, such as the redshift, the bending of angles and the observable flux of radiation as a function of time, see for example~\cite{wAkT68,jJ69,kLrR79,hLvF06,vFkKhL07} and references therein.

Recently, it was argued in~\cite{lKdMcB13} that a significant fraction of photons which are emitted from particles inside a collapsing object might be detected by distant observers as blueshifted rather than redshifted. This effect, which was explicitly verified for radial light rays traversing a spherical, marginally bound dust cloud, is due to the fact that the photons are blueshifted inside the collapsing region and that, for some of these photons, this blueshift can actually be larger than the redshift in the outside region. The spectrum of the radiation emitted by a collapsing spherical dust cloud for a simple emissivity function has also been computed in~\cite{lKdMcB13}, and in~\cite{lKdMcB13b} the possibility of using this spectrum in order to distinguish the formation of a black hole from the formation of a naked singularity was discussed.

In this article, we consider a related but different scenario, in which photons from a distant emitter traverse a collapsing object without interacting with its matter content and are detected by a distant observer. As an example, we may think of radiation emitted from a bright electromagnetic source such as a distant galaxy or supernova and which crosses a collapsing dust cloud before it is detected on earth. For sufficiently large wavelengths the photons do not interact significantly with the dust cloud. Such photons are only affected by the gravitational field of the collapsing object, and thus they may provide direct information on its mass distribution. Here, we focus our attention on the frequency shift of such photons, assuming that both the emitter and the observer are located very far from the object and are stationary. Note that for \emph{stationary} objects, such as stars in equilibrium, there is no frequency shift in this scenario since the photon's energy is conserved along its trajectory. However, as we show, for the case of a \emph{collapsing} object, the dynamics of the spacetime implies a nontrivial frequency shift of the photons, which, for the simplest case of spherical symmetry and pressureless matter satisfying weak assumptions is always towards the \emph{red}.

In order to model this problem, we consider the idealized situation of an isolated object which collapses in an asymptotically flat vacuum exterior spacetime. Here, we are thinking about astrophysical situations in which both the light source and the observer are located at distances which are large compared to the Schwarzschild radius of the object but small on cosmological scales. Therefore, we study light rays sent by an asymptotic emitter and detected by an asymptotic observer described by null geodesics propagating from past to future null infinity. In this work, we compute the frequency shift of photons traveling on such geodesics for the simple case of a spherical dust cloud and radial geodesics traversing the cloud through its center.

We start in Sec.~\ref{Sec:Hom} with the case of a homogeneous density distribution, for which the interior spacetime is modeled by a (contracting) Friedmann-Robertson-Walker (FRW) spacetime. When traveling from past null infinity towards the cloud, the photons are blueshifted since they approach a region of stronger gravity. In the interior region the same photons are also blueshifted as they propagate inside a collapsing universe. However, we show by explicit calculation that this blueshift is overcompensated by redshift effects, so that the total frequency shift is always a \emph{redshift}. These effects consist of the gravitational redshift experienced by the photons when they travel from the surface of the cloud to future null infinity and to Doppler effects due to the motion of the surface of the cloud with respect to static observers. In this sense, photons lose energy each time they cross a collapsing dust cloud. In Sec.~\ref{Sec:Hom} we also discuss the dependency of this redshift effect on the compactness ratio of the cloud at the moment it is penetrated by the light ray, and we derive bounds on the blueshift effect discussed in~\cite{lKdMcB13}. In contrast to the work in~\cite{lKdMcB13} we do not restrict ourselves to the marginally bound case.

Next, in Sec.~\ref{Sec:Relation} we derive a geometric identity, valid for any spherically symmetric spacetime, which provides an interesting relation between the redshift factor along in- and outgoing radial null rays, the velocity field of the collapsing shells, and the radial pressure. For the particular case of dust collapse, the radial pressure vanishes and the identity yields a direct relation between the redshift factor and the compactness ratio of the shells. Based on this identity, we derive in Sec.~\ref{Sec:TB} an explicit upper bound on the frequency shift for photons with zero angular momentum which travel from past to future null infinity through the center of a spherical, collapsing, inhomogenous dust cloud. We derive this bound first in the marginally bound case and then in the bounded case, assuming only a positive mass density, the absence of shell-crossing singularities, and assuming that the total energy of the shells is a nonincreasing function of the areal radius. The importance of our bound is that it reveals that such photons are always redshifted with respect to asymptotic stationary observers.

In Sec.~\ref{Sec:Num} we compute the total frequency shift by numerical means for a particular family of time-symmetric inhomogeneous initial data for the collapsing dust cloud. This family is described by the initial density profile of the cloud which is parametrized by the central density, the initial radius of the cloud, and an additional parameter which measures how flat the density profile is. By choosing this last parameter very large, we recover the homogeneous case discussed in Sec.~\ref{Sec:Hom} to arbitrary accuracy. In this way, we can verify the validity of our code by comparing the frequency shift computed numerically with the results from Sec.~\ref{Sec:Hom}. We also show that the upper bound for the frequency shift derived in Sec.~\ref{Sec:TB} is consistent with our numerical results. Then, we focus our attention on a particular subfamily with constant total mass and initial radius, and we compare the total redshift measured by a distant observer as a function of its proper time. We show that this function depends on the inner mass distribution of the collapsing cloud, a fact that could in principle be exploited to infer the density profile of the cloud from the measurement of the redshift function. Finally, conclusions are drawn in Sec.~\ref{Sec:Conclusions}, where we also discuss implication of the total redshift effect to the weak cosmic censorship conjecture and potential applications to cosmological scenarios.

Throughout this work we use geometrized units, where the gravitational constant and the speed of light are set equal to one. Vector and tensor fields are denoted by bold face symbols, for instance ${\bf u} = u^{\mu}\partial_{\mu}$. The application of a vector field ${\bf u}$ on a function $f$ is written as ${\bf u}[f] = u^{\mu}\partial_{\mu}f$.

\section{The homogeneous case}
\label{Sec:Hom}

\begin{figure}[h!]
\begin{center}
\includegraphics[width=9.0cm]{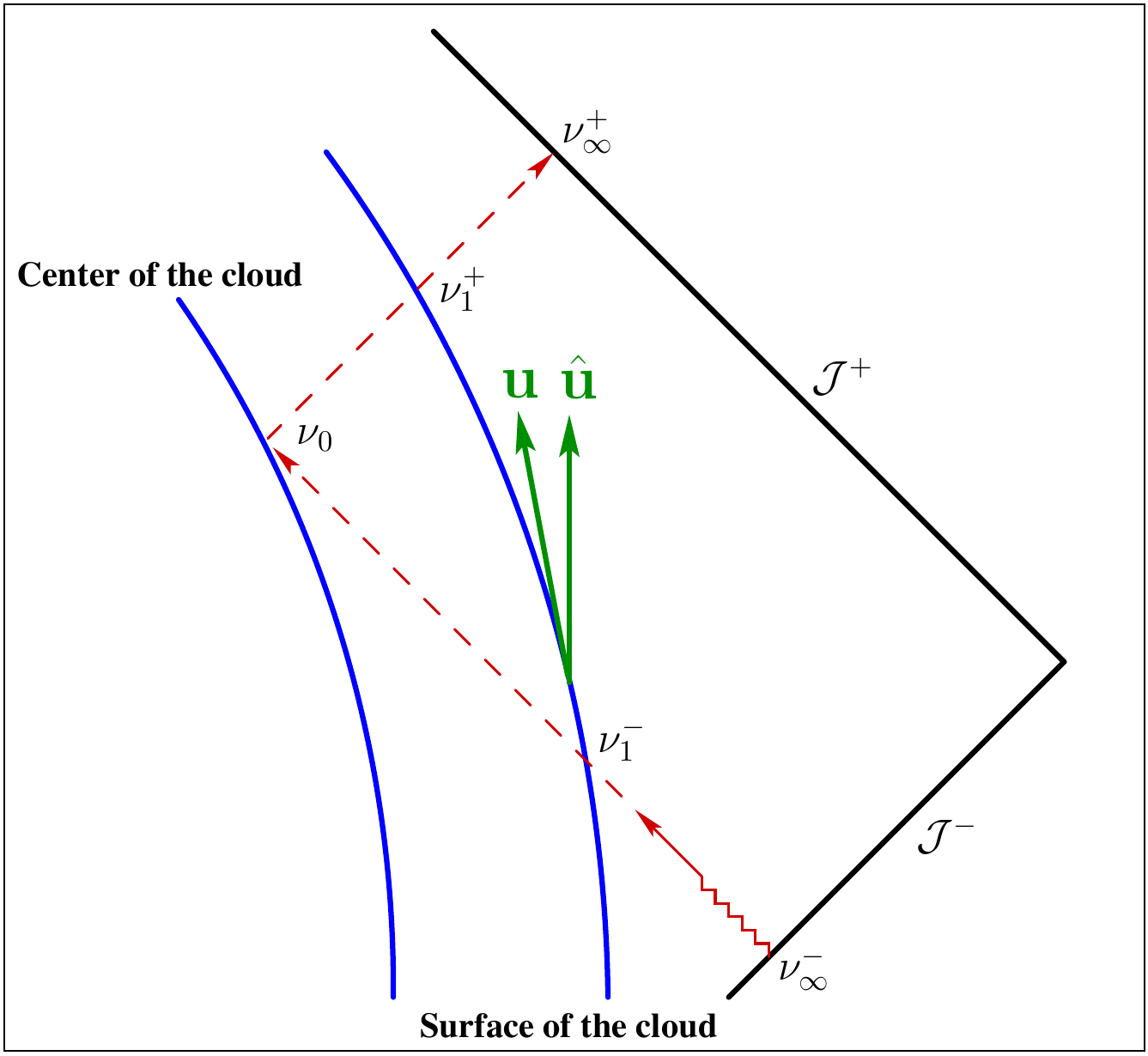}
\end{center}
\caption{\label{Fig:Cartoon} Conformal diagram illustrating the path of a photon traveling from past $(\Scri^-)$ to future null infinity $(\Scri^+)$. The $\nu$'s refer to the frequencies of the photons at the indicated events. Also shown are the four-velocity of stationary ($\hat{\bf u}$) and free falling observers (${\bf u}$) at the surface of the cloud.}
\end{figure}

In this section, we derive the total redshift effect in the simplest case of a spherical, homogeneous collapsing dust cloud of finite radius. The interior spacetime corresponds to a contracting FRW universe and, as a consequence of our assumptions and Birkhoff's theorem, the exterior is described by a Schwarzschild spacetime of mass $m_1$. The matching is performed by identifying a co-moving sphere containing mass $m_1$ and of areal radius $r_1(\tau)$ in the FRW spacetime with the sphere of equal areal radius in Schwarzschild spacetime whose world sheet is spanned by freely falling radial geodesics. Consequently, the usual matching conditions are automatically satisfied~\cite{MTW-Book}. We consider a photon with zero angular momentum traveling from past $(\Scri^-)$ to future null infinity $(\Scri^+)$ through the center of the cloud, see Fig.~\ref{Fig:Cartoon}. The frequency shift of this photon has three different contributions. The first one consists of a combined effect of a gravitational blueshift from $\Scri^-$ to the surface of the cloud with a Doppler redshift due to the motion of the surface relative to static observers. The second contribution consists of a blueshift due to the contraction of the universe. The third contribution is similar to the first one but in this case both the Doppler and the gravitational effect result in a redshift.
Let us now explicitly compute these three contributions. The gravitational blueshift measured by static observers when the photon travels from $\Scri^-$ to the surface of the cloud is given by $\left(1 - 2m_1/r_1^-\right)^{-1/2}$, where $m_1$ is the total mass of the cloud and $r_1^-$ its areal radius at the time the photon penetrates it. The Doppler shift due to the motion of the surface is given by ${\bf g}\left({\bf u},{\bf k}_-\right) / {\bf g}\left(\hat{\bf u},{\bf k}_-\right)$, where ${\bf g}$ is the spacetime metric, ${\bf u}$ and $\hat{\bf u}$ are, respectively, the four-velocities of co-moving and static observers at the surface of the cloud, and ${\bf k}_-$ is a radial incoming null vector, proportional to the four-momentum of the photon. In terms of the unit outgoing radial vector field ${\bf w}$ orthogonal to ${\bf u}$, we define ${\bf k}_- = {\bf u} - {\bf w}$. In order to determine the vector field $\hat{\bf u}$, we first note that the motion of the surface of the cloud is given by the free fall equation
\begin{equation}\label{Eq:free_fall}
\frac{1}{2}\dot{r}^2 - \frac{m_1}{r} = E_1,
\end{equation}
with $E_1$ the total energy of the trajectory and $\dot{r} := {\bf u}[r]$. For the following, we assume that $-1/2 < E_1 \leq 0$, which means that the trajectory is bounded $(E_1 < 0)$ or marginally bound $(E_1 = 0)$ and that $2m_1/r < 1$ in the region where $\dot{r}^2$ is small, so that there are no trapped surfaces lying inside this region. Combining Eq.~(\ref{Eq:free_fall}) with the requirement that $\hat{\bf u}[r] = 0$ and the invariant definition of the Misner-Sharp mass function~\cite{cMdS64} $m(r) = r[1 - {\bf g}(dr,dr)]/2 = r(1 + {\bf u}[r]^2 - {\bf w}[r]^2)/2$, which yields ${\bf w}[r] = \sqrt{1 + 2E_1}$, we obtain
\begin{equation}
\hat{\bf u} = \frac{1}{\sqrt{1 - \frac{2m_1}{r}}} \left( \sqrt{1 + 2E_1} {\bf u} - \dot{r}{\bf w} \right).
\end{equation}
Therefore, the Doppler shift is
\begin{equation}
\frac{{\bf g}\left({\bf u},{\bf k}_-\right)}{{\bf g}\left(\hat{\bf u},{\bf k}_-\right)} 
 = \left. \frac{\sqrt{1 - \frac{2m_1}{r}}}{\sqrt{1 + 2E_1} - \dot{r}} \right|_{r = r_1^-},
\end{equation}
and the frequency shift between a static observer at $\Scri^-$ and a free falling observer at the surface of the cloud is
\begin{equation}\label{Eq:Redshift_past}
\frac{\nu_1^-}{\nu_\infty^-} = \frac{1}{\sqrt{1 + 2E_1} + \sqrt{{\cal C}_1^- + 2E_1}}.
\end{equation}
Here and in the following, ${\cal C} := 2m/r$ denotes the compactness ratio. Note that in the particular case of marginally bound collapse, for which $E_1 = 0$, the frequency shift simplifies to $ \nu_1^- /\nu_\infty^- =  \left(1 + \sqrt{{\cal C}_1^-}\right)^{-1/2} < 1$, which shows that in this case the Doppler redshift dominates the gravitational blueshift.

In the interior of the collapsing cloud, the photon experiences a blueshift due to the contraction of the FRW universe given by (see, for instance,~\cite{Straumann-Book})
\begin{equation}
\frac{\nu_1^+}{\nu_1^-} 
 = \frac{a(\tau_1^-)}{a(\tau_1^+)} = \frac{{\cal C}_1^+}{{\cal C}_1^-},
\end{equation}
where we have used the fact that the scale factor $a(\tau)$ is proportional to the areal radius $r$ along the surface of the cloud, and $\tau$ denotes proper time along a radial observer co-moving with the surface.

Finally, as a calculation similar to the one leading to Eq.~(\ref{Eq:Redshift_past}) shows, the redshift between the surface of the cloud and a static observer at $\Scri^+$ is given by
\begin{equation}
\frac{\nu_\infty^+}{\nu_1^+} = \sqrt{1 + 2E_1} - \sqrt{{\cal C}_1^+ + 2E_1} < 1,
\end{equation}

\subsection{The total redshift}

From the results so far, we conclude that the total frequency shift experienced by the photon is
\begin{equation}\label{Eq:TotalRedShift}
\frac{\nu_\infty^+}{\nu_\infty^-} = \frac{{\cal C}_1^+}{{\cal C}_1^-} \frac{\sqrt{1 + 2E_1} - \sqrt{{\cal C}_1^+ + 2E_1} }{\sqrt{1 + 2E_1} + \sqrt{{\cal C}_1^- + 2E_1}}.
\end{equation}
This formula can be simplified by finding the explicit relation between the compactness ratios ${\cal C}_1^+$ and ${\cal C}_1^-$. For this, we integrate the light rays inside the cloud. In order to do so, we first note that the areal radius is proportional to the scale factor, and thus $r = r_0 a(\tau)$ for some constant $r_0$ independent of $\tau$. Substituting this relation into Eq.~(\ref{Eq:free_fall}) one obtains the Friedmann equation
\begin{equation}\label{Eq:Friedmann}
\dot{a}^2 = \frac{\lambda}{a} - k,
\end{equation}
with $\lambda := 2m/r_0^3$ and the spatial curvature parameter $k := -2E_1/r_0^2$. Following the usual convention, we choose $r_0$ such that $k=1$ when $E_1 < 0$ in which case the spatial geometry is $S^3$. In the marginally bound case, $k$ is zero, corresponding to a flat spatial geometry and $r_0$ is undetermined. The FRW spacetime metric is
\begin{equation}
{\bf g} = -d\tau^2 + a(\tau)^2 \left( dR^2 + \sinn^2(R) d\Omega^2  \right),
\end{equation}
with $\sinn(R) = \sin(R)$ for $k=1$ and $\sinn(R) = R$ for $k=0$, and it follows that $r_0 = \sinn(R)$. Integrating along the radial null ray inside the cloud gives
\begin{equation}
2R_1 = \int_{\tau_1^-}^{\tau_1^+} \frac{d\tau}{a(\tau)}.
\end{equation}
Using the Friedmann equation~(\ref{Eq:Friedmann}) yields
\begin{equation}
R_1 = \left\{ \begin{array}{ll} 
\sqrt{\frac{a(\tau_1^-)}{\lambda}} - \sqrt{\frac{a(\tau_1^+)}{\lambda}},  & k = 0, \\
\arccos\left( \sqrt{\frac{a(\tau_1^+)}{\lambda}} \right) 
- \arccos\left( \sqrt{\frac{a(\tau_1^-)}{\lambda}} \right), & k = 1.
\end{array} \right.
\end{equation}
This leads to the following relation between the compactness ratios ${\cal C}_1^+$ and ${\cal C}_1^-$:
\begin{equation}
\sqrt{\frac{{\cal C}_1^-}{{\cal C}_1^+}} = \sqrt{1 + 2E_1} - \sqrt{{\cal C}_1^- + 2E_1}
\quad\hbox{or}\quad
\sqrt{\frac{{\cal C}_1^+}{{\cal C}_1^-}} = \sqrt{1 + 2E_1} + \sqrt{{\cal C}_1^+ + 2E_1}.
\end{equation}
Using these relations to eliminate ${\cal C}_1^+$ in Eq.~(\ref{Eq:TotalRedShift}) yields, finally,
\begin{equation}\label{Eq:FinalRedShift}
\frac{\nu_\infty^+}{\nu_\infty^-} = \frac{1}{1 - {\cal C}_1^-}\left[
 1 - \frac{{\cal C}_1^-}{\left( \sqrt{1 + 2E_1} - \sqrt{{\cal C}_1^- + 2E_1} \right)^2} \right].
\end{equation}
Since we are interested in the collapsing case, for which $\dot{r} < 0$, and since we assume that the light ray lies outside the event horizon, such that ${\cal C}_1^+ < 1$, the compactness ratio ${\cal C}_1^-$ at the moment of penetration is restricted to the interval
\begin{equation}\label{Eq:C1Interval}
-2E_1 < {\cal C}_1^- < \frac{1}{4(1 + 2E_1)}.
\end{equation}
This in turn restricts $E_1$ to the interval $-1/4 < E_1 \leq 0$.

When the light ray penetrates the cloud at the instant it is at rest $(\dot{r} = 0)$, ${\cal C}_1^- = -2E_1 > 0$, and the total frequency shift is
\begin{equation}
\frac{\nu_\infty^+}{\nu_\infty^-} = \frac{1 + 4E_1}{1 + 4E_1 + 4E_1^2} < 1,
\end{equation}
implying a redshift. When ${\cal C}_1^-$ approaches its upper value $[4(1+2E_1)]^{-1}$, the frequency ratio $\nu_\infty^+/\nu_\infty^-$ converges to zero. This is expected since in this case the light ray exits the cloud at a point which lies very close to the event horizon, implying a large redshift in the Schwarzschild region. In order to analyze the behavior of the frequency shift inside the interval~(\ref{Eq:C1Interval}) it is convenient to introduce the dimensionless quantities $x:=-\dot{r}/\sqrt{1 + 2E}$ and $\varepsilon_1:=-2E_1/(1+2E_1)$, $0 < \varepsilon_1 < 1$, in terms of which Eq.~(\ref{Eq:FinalRedShift}) can be rewritten as
\begin{equation}\label{Eq:FinalRedShiftBis}
\frac{\nu_\infty^+}{\nu_\infty^-} = (1+\varepsilon_1)
\frac{1-\varepsilon_1 - 2x_1^-}{(1+x_1^-)(1-x_1^-)^3}
\end{equation}
with $0 < x_1^- < (1-\varepsilon_1)/2$. By explicitly differentiating the right-hand side with respect to $x_1^-$ it is not difficult to prove that the frequency ratio $\nu_\infty^+/\nu_\infty^-$ is a monotonously decaying function of $x_1^-$ inside the interval of interest $0 < x_1^- < (1-\varepsilon_1)/2$, so that this ratio is always smaller than one. This leads to our first important conclusion: \emph{a radial light ray traversing a collapsing, homogeneous dust cloud through its center is always redshifted}.

In the limit $\varepsilon_1\to 0$, corresponding to the marginally bound case, Eq.~(\ref{Eq:FinalRedShiftBis}) simplifies to
\begin{equation}\label{Eq:FinalRedShiftBisExp}
\frac{\nu_\infty^+}{\nu_\infty^-} = \frac{1-2x_1^-}{(1+x_1^-)(1-x_1^-)^3}
 = 1 - 2(x_1^-)^3 + {\cal O}^4(x_1^-),
\end{equation}
and the redshift factor $z := \nu_\infty^-/\nu_\infty^+ - 1 = 2(x_1^-)^3 + {\cal O}^4(x_1^-)$ scales like the compactness ratio to the power of $3/2$ for small $x_1^-$.

\subsection{On the blueshift of photons emitted from inside the cloud}

As mentioned in the introduction, in recent work~\cite{lKdMcB13} it has been shown that photons emitted from dust particles inside the cloud may be blue- instead of redshifted when observed by a distant, stationary observer. Here we analyze this effect for a homogeneous dust cloud. The particular case of marginally bound collapse has already been analyzed in~\cite{lKdMcB13}; here we generalize the discussion to the more generic bounded collapse.

Since photons are blueshifted inside the cloud, an upper bound for the frequency ratio of photons emitted from the dust particles is provided by the ratio corresponding to those photons which are emitted from the surface of the cloud, travel inwards through the center of the cloud and escape to $\Scri^+$. From the results derived above we easily obtain the following expression for the frequency shift of such photons:
\begin{equation}\label{Eq:BlueShift}
\frac{\nu_\infty^+}{\nu_1^-} = \sqrt{1+\varepsilon_1}
\frac{1-\varepsilon_1 - 2x_1^-}{(1-x_1^-)^3}.
\end{equation}
In the marginally bound case, corresponding to the Oppenheimer-Snyder model~\cite{jOhS39}, this expression simplifies to $\nu_\infty^+/\nu_1^- = (1-2x_1^-)/(1-x_1^-)^3 = 1 + x_1^- + {\cal O}^2(x_1^-)$, showing that a blueshift is possible for small enough $x_1^-$.

In general, the expression on the right-hand side of Eq.~(\ref{Eq:BlueShift}) has a local maximum at $x_1^- = (1 - 3\varepsilon_1)/4$, where
$$
\left. \frac{\nu_\infty^+}{\nu_1^-} \right|_{max} 
 = \frac{32}{27}\frac{1}{(1 + \varepsilon_1)^{3/2}}.
$$
Therefore, photons which emanate from the surface of the cloud with $x_1^- \simeq (1-3\varepsilon_1)/4$ are blueshifted provided that $\varepsilon_1 < \varepsilon_1^* := 8\cdot 2^{1/3}/9 - 1 \simeq 0.1199$. Above this threshold the frequency shift is always smaller than one. Therefore, photons with zero angular momentum which are emitted by dust particles in a bounded, homogeneous dust collapse for which $\varepsilon_1 > \varepsilon_1^*$ are always redshifted when observed by a distant stationary observer.

\section{Relation between the compactness ratio and the redshift factor}
\label{Sec:Relation}

In this section, we derive a geometric identity which is valid for any spherically symmetric spacetime $(M,{\bf g})$ satisfying Einstein's field equations with the stress-energy tensor ${\bf T}$ of a fluid. Such a spacetime has the form $M = \tilde{M}\times S^2$ with metric ${\bf g} = \tilde{\bf g} + r^2(d\vartheta^2 + \sin^2\vartheta d\varphi^2)$, where $\tilde{\bf g} = \tilde{g}_{ab} dx^a dx^b$ is a two-dimensional Lorentzian metric on $\tilde{M}$ and $r$ is a positive function on $\tilde{M}$ describing the areal radius of the invariant two sphere $\{ p \}\times S^2$ for each $p\in \tilde{M}$. The stress-energy tensor has the form
\begin{equation}
{\bf T} = (\rho u_a u_b + p_\perp w_a w_b) dx^a dx^b
 + r^2 p_{||}(d\vartheta^2 + \sin^2\vartheta d\varphi^2),
\label{Eq:StressEnergyT}
\end{equation}
with ${\bf u} = u^a\partial_a$ the four-velocity of the fluid and ${\bf w} = w^a\partial_a$ the outward radial unit vector orthogonal to ${\bf u}$. Here, $\rho$, $p_\perp$ and $p_{||}$ denotes the energy density, radial and tangential pressure of the fluid. Einstein's field equations imply (see, for instance, Ref.~\cite{eCnOoS13})
\begin{eqnarray}
(\tilde{\nabla}_a\tilde{\nabla}_b r)^{tf} &=& -2\pi r(\rho + p_\perp)(u_a u_b + w_a w_b) ,
\label{Eq:Einstein1}\\
\tilde{\Delta} r &=& \frac{2m}{r^2} - 4\pi r(\rho - p_\perp),
\label{Eq:Einstein2}
\end{eqnarray}
where $\tilde{\nabla}$ is the covariant derivative associated with $(\tilde{M},\tilde{\bf g})$, $\tilde{\Delta} r := \tilde{g}^{ab}\tilde{\nabla}_a\tilde{\nabla}_b r$, and $(tf)$ denotes the trace-free part.

The vector fields ${\bf u}$ and ${\bf w}$ give rise to the in- and outgoing radial null vector fields ${\bf k}_- := {\bf u} - {\bf w}$ and ${\bf k}_+ := {\bf u} + {\bf w}$. Although they are tangent to null geodesics, their integral curves are not necessarily affinely parametrized. For the following, we introduce the quantities $\beta_\pm$, defined by
\begin{equation}
\nabla_{\bf k_\pm}{\bf k_\pm} = -\beta_\pm{\bf k_\pm},
\label{Eq:BetaDef}
\end{equation}
which determine the frequency shift along in- and outgoing radial null geodesics as measured by co-moving observers with four-velocity ${\bf u}$. Specifically, if $\nu$ is the frequency of a radial light signal measured by an observer with four-velocity ${\bf u}$, then $\nu$ satisfies the differential equation
\begin{displaymath}
{\bf k}_\pm[\log\nu] = \beta_\pm
\end{displaymath}
along incoming (-) and outgoing (+) null geodesics. Using the fact that ${\bf k}_\pm$ are radial and null, and using their relative normalization ${\bf g}({\bf k}_+,{\bf k}_-) = -2$, it is not difficult to see that
\begin{equation}
\nabla_{\bf k_\pm}{\bf k_\mp} = \beta_\pm{\bf k_\mp},
\label{Eq:Beta}
\end{equation}
and that the acceleration of the fluid elements satisfies
\begin{equation}\label{Eq:acceleration}
{\bf a} := \nabla_{\bf u} {\bf u} = -\frac{\beta_+ - \beta_-}{2} {\bf w}.
\end{equation}

Next, we introduce the (absolute value of the) velocity of the free falling fluid elements: $V := -{\bf u}[r]$, where the minus sign indicates that the fluid elements are moving towards the center. Our key identity is obtained by applying the null vector fields ${\bf k}_\pm$ on $V$. Using the fact that $2{\bf u} = {\bf k}_+ + {\bf k}_-$ and Eqs.~(\ref{Eq:BetaDef},\ref{Eq:Beta}), we obtain
\begin{equation}
{\bf k}_\pm[V] = -{\bf k}_\pm{\bf u}[r]
 = -\frac{1}{2}\left[ k_\pm^a k_\pm^b(\tilde{\nabla}_a\tilde{\nabla}_b r)^{tf}
 - \tilde{\Delta} r \right] \pm \beta_\pm {\bf w}[r].
\label{Eq:Id2}
\end{equation}
At this point we use Einstein's field equations~(\ref{Eq:Einstein1},\ref{Eq:Einstein2}) in order to eliminate the second derivatives of $r$ and obtain from Eq.~(\ref{Eq:Id2}) the identity
\begin{equation}\label{Eq:Key}
{\bf k}_\pm[V] = 4\pi r p_\perp + \frac{m}{r^2} \pm \beta_\pm {\bf w}[r].
\end{equation}
In order to explain the relevance of this identity, we restrict ourselves to the particular case of marginally bound dust collapse, in which case $p_\perp = 0$, ${\bf a} = 0$, ${\bf w}[r] = 1$, and $V = -{\bf u}[r] = \sqrt{2m/r}$ is the square root of the compactness ratio, due to the fact that each collapsing shell has zero total energy. Under these assumptions Eq.~(\ref{Eq:Key}) simplifies considerably and yields
\begin{equation}\label{Eq:KeyMB}
{\bf k}_\pm[\log\nu] = \beta_\pm = \pm {\bf k}_{\pm}[V] \mp \frac{m}{r^2}.
\end{equation}
This is our key identity which gives a direct relation between the frequency shift along radial null rays and the compactness ratio. Consequences of this identity and its generalization to the bounded dust collapse case will be discussed in the next section.

\section{The total redshift effect for the inhomogeneous collapse}
\label{Sec:TB}

The spherically symmetric solutions of the field equations describing a self-gravitating dust configuration can be explicitly parametrized in terms of co-moving, synchronous coordinates $(\tau,R,\vartheta,\varphi)$. Here, $R=const.$ describes the world sheet of the collapsing dust shells, where the label $R$ is chosen such that it coincides with the shells' areal radius at initial time $\tau=0$. $\tau$ is the proper time measured by a radial observer moving along a collapsing dust shell, and $(\vartheta,\varphi)$ are standard polar coordinates on the invariant two-spheres. The metric is determined by the function $r(\tau,R)$ which describes the areal radius at the event $(\tau,R,\vartheta,\varphi)$. Therefore, for fixed $R$, the function $\tau\mapsto r(\tau,R)$ describes the evolution of the dust shell labeled by $R$, and according to the definition of $R$, $r(0,R) = R$. This function is determined by the free fall equation~(\ref{Eq:free_fall}) with $m_1$ replaced with $m(R)$, the (conserved) mass contained inside the dust shell $R$, and $E_1$ replaced with $E(R)$, the total energy of the dust shell $R$.

In terms of the coordinates $(\tau,R)$ the radial metric $\tilde{\bf g}$, the four-velocity ${\bf u}$, and the vector field ${\bf w}$ are (see~\cite{MTW-Book} and references therein)
\begin{equation}
\tilde{\bf g} = -d\tau^2 + \frac{r'(\tau,R)^2}{1+2E(R)} dR^2,\qquad
{\bf u} = \frac{\partial}{\partial\tau},\quad
{\bf w} = \frac{\sqrt{1 + 2E(R)}}{r'(\tau,R)} \frac{\partial}{\partial R},
\label{Eq:MetricSol}
\end{equation}
where the prime denotes partial differentiation with respect to $R$. The causal structure of the resulting spacetime has been studied extensively, see for example~\cite{dC84,rN86,Joshi-Book,nOoS11}, and the final state of the collapse may be a black hole or a naked singularity. Here, we focus mainly on the black hole case.

In the remaining of this section, we prove the total redshift effect under the following assumptions:
\begin{enumerate}
\item[(i)] The mass density is nonnegative, such that $m\geq 0$ and $m'\geq 0$.
\item[(ii)] $E'\leq 0$.
\item[(iii)] $r' > 0$, so that there are no shell-crossing singularities.
\end{enumerate}
See~\cite{rN86} for conditions on the functions $m(R)$ and $E(R)$ characterizing the collapse which guarantee the fulfillment of assumption (iii).

\subsection{The marginally bound case}

As explained in the previous section, the particular case of marginally bound collapse is characterized by zero total energy for all the dust shells, $E(R) = 0$, and in this case we have the simplified identity~(\ref{Eq:KeyMB}) which relates the frequency change with the change of $x:= V = \sqrt{2m/r}$ along in- and outgoing radial null geodesics. In the outgoing case, integration of Eq.~(\ref{Eq:KeyMB}) from the center to the surface of the cloud (see Fig.~\ref{Fig:Cartoon}) yields the following frequency shift:
\begin{equation}\label{Eq:nu1/nu0}
\frac{\nu_1^+}{\nu_0} = \exp\left( x_1^+ - \int_0^{R_1}
\frac{m}{r^2} r' dR \right),
\end{equation}
where we have parametrized the light ray by the coordinate $R$, used the fact that ${\bf k}_+[R] = 1/r'$, and the fact that $x\to 0$ as $R\to 0$ along the light ray. Similarly in the ingoing case, integration of Eq.~(\ref{Eq:KeyMB}) from the surface of the cloud to its center yields
\begin{equation}\label{Eq:nu0/nu1}
\frac{\nu_0}{\nu_1^-} = \exp\left( x_1^- + \int_0^{R_1}
\frac{m}{r^2} r' dR \right),
\end{equation}
which, in contrast to $\nu_1^+/\nu_0$ is manifestly larger than one, implying a blueshift. Therefore, the frequency shift of photons emitted from the surface of the cloud towards its center and received by a co-moving observer at a diametrically opposite point on the surface of the cloud is given by
\begin{equation}\label{Eq:nu1/nu1}
\frac{\nu_1^+}{\nu_1^-} = \exp\left[ x_1^- + x_1^+ + \int_0^{R_1}
m(R)\left(  \left.\frac{r'}{r^2}\right|_- - \left.\frac{r'}{r^2}\right|_+ \right) dR \right],
\end{equation}
where the subscripts $-$ and $+$ indicate evaluation along the in- and outgoing null rays, respectively.

Next, we add to this the contributions from the frequency shift between a static observer at $\Scri^-$ and a free falling observer at the surface of the cloud, and the one from the redshift between an observer at the surface of the cloud and a static observer at $\Scri^+$. These contributions are independent of the interior of the cloud and have been discussed in detail in Sec.~\ref{Sec:Hom}. The result for the total frequency shift for a radial light ray extending from $\Scri^-$ to $\Scri^+$ is
\begin{equation}\label{Eq:nu_infty/nu_infty}
\frac{\nu_{\infty}^+}{\nu_{\infty}^-} = \frac{1 - x_1^+}{1+x_1^-} \exp\left[ x_1^+ + x_1^- + \int_0^{R_1} m(R)\left(  \left.\frac{r'}{r^2}\right|_- - \left.\frac{r'}{r^2}\right|_+ \right) dR  \right].
\end{equation}
In order to find an upper bound for this frequency shift, we first note that along radial null rays
$$
\frac{d}{dR}\frac{1}{r} = \pm r' {\bf k}_\pm\left[ \frac{1}{r} \right]
 = -\frac{r'}{r^2} \mp r'\frac{\dot{r}}{r^2},
$$
so that we can rewrite the integral on the right-hand side of Eq.~(\ref{Eq:nu_infty/nu_infty}) as
$$
-\int_0^{R_1} m(R)\left(  \left.\frac{d}{dR}\frac{1}{r}\right|_- - \left.\frac{d}{dR}\frac{1}{r}\right|_+ \right)dR + \int_0^{R_1} m(R)\left( \left.\frac{r' \dot{r}}{r^2}\right|_+ + \left.\frac{r' \dot{r}}{r^2}\right|_- \right) dR.
$$
Using integration by parts in the first term leads to the following result:
\begin{equation}\label{Eq:TotalRedShiftMB}
\frac{\nu_{\infty}^+}{\nu_{\infty}^-} = f_+(x_1^+) f_-(x_1^-)
\exp\left[ \int_0^{R_1} m'(R)\left( \left.\frac{1}{r}\right|_- - \left.\frac{1}{r}\right|_+ \right) + m(R)\left( \left.\frac{r' \dot{r}}{r^2}\right|_+ + \left.\frac{r' \dot{r}}{r^2}\right|_- \right) dR  \right],
\end{equation}
with the functions $f_\pm$ defined as
\begin{equation}\label{Eq:fpm}
f_+(x) := (1-x)\exp(x + x^2/2),\qquad
f_-(x) := (1+x)^{-1}\exp(x - x^2/2),\quad, x\geq 0.
\end{equation}
Next, we claim that the expression inside the square parenthesis on the right-hand side of Eq.~(\ref{Eq:TotalRedShiftMB}) is negative or zero. In order to see this, we first note that the functions $m$ and $m'$ are nonnegative because we assume a nonnegative mass density. Next, since we assume a collapsing cloud, it is evident that for a given mass shell parametrized by a fixed value of $R$, its areal radius $r$ is larger at the moment it is penetrated by the light ray than at the moment when the light ray exits it, so that $r^- > r^+$. Finally, we use again the fact that $\dot{r} < 0$, and the assumption that there are no shell-crossing singularities, which is guaranteed by $r' > 0$. With these observations in mind we conclude that the total frequency shift is bounded from above according to
\begin{equation}\label{Eq:bound}
\frac{\nu_{\infty}^+}{\nu_{\infty}^-} \leq f_+(x_1^+) f_-(x_1^-),
\end{equation}
with the functions $f_\pm$ defined in Eq.~(\ref{Eq:fpm}). It is not difficult to verify that these functions are monotonically decreasing and satisfy $f_+(0) = f_-(0) = 1$, with $f_\pm(x)\leq 1$ for all $x\geq 0$. \emph{This proves that the total frequency shift is towards the red, as in the homogeneous case}.

Since $x_1^+ \geq x_1^-$ and $f_+$ is monotonously decreasing, we can further estimate $f_+(x_1^+) \leq f_+(x_1^-)$ and obtain the following bound in terms of $x_1^-$ only:
\begin{equation}
\frac{\nu_{\infty}^+}{\nu_{\infty}^-} \leq f_+(x_1^-) f_-(x_1^-) = \frac{1 - x_1^-}{1 + x_1^-}\exp(2x_1^-).
\end{equation}
For small values of $x_1^-$ we obtain, in particular,
\begin{equation}
\frac{\nu_{\infty}^+}{\nu_{\infty}^-} \leq 1 - \frac{2}{3}\left(x_1^-\right)^3  
 + {\cal O}^5(x_1^-).
\end{equation}
Comparison with the result from the homogeneous case in Eq.~(\ref{Eq:FinalRedShiftBisExp}) reveals the same dependency of $({\cal C}_1^-)^{3/2}$ of the redshift factor $z = \nu_{\infty}^- / \nu_{\infty}^+ - 1$ in terms of the compactness ratio ${\cal C}_1^- = (x_1^-)^2$, but with the constant $-2/3$ instead of $-2$. Therefore, for small ${\cal C}_1^-$, our upper bound reproduces correctly the scaling in terms of the compactness ratio, but with a larger constant.

\subsection{The bounded case}
\label{Sec:Generic_case}

Next, we generalize the bound~(\ref{Eq:bound}) to the case where each dust shell has zero or negative total energy, $E(R) \leq 0$. In this case, we follow the same lines of arguments as in the marginally bound case starting from the identity~(\ref{Eq:Key}) with $p_\perp = 0$ and ${\bf w}[r] = \sqrt{1+2E}$. Introducing the dimensionless variable $x := V/\sqrt{1+2E} = \sqrt{\frac{2m}{r} + 2E}/\sqrt{1+2E}$, as in Sec.~\ref{Sec:Hom}, and integrating Eq.~(\ref{Eq:Key}) along in- and outgoing radial light rays, one finds the generalization of Eqs.~(\ref{Eq:nu1/nu0})~and~(\ref{Eq:nu0/nu1}), 
\begin{eqnarray}
\frac{\nu_1^+}{\nu_0} &=& \exp\left( x_1^+ - \int_0^{R_1} \frac{m}{1+2E} \frac{r'}{r^2} dR + \int_0^{R_1} \frac{x E'}{1+2E} dR\right),\\
\frac{\nu_0}{\nu_1^-} &=& \exp\left( x_1^- + \int_0^{R_1} \frac{m}{1+2E} \frac{r'}{r^2} dR + \int_0^{R_1} \frac{x E'}{1+2E} dR\right),
\end{eqnarray}
where we have used the fact that ${\bf k}_\pm[R] = \pm\sqrt{1+2E}/r'$. This leads to the following generalization of Eq.~(\ref{Eq:nu_infty/nu_infty}) for the total frequency shift:
\begin{equation}\label{Eq:nu_infty/nu_infty_generic}
\frac{\nu_{\infty}^+}{\nu_{\infty}^-} = \frac{1 - x_1^+}{1+x_1^-} 
\exp\left[ x_1^+ + x_1^-  + \int_0^{R_1} \frac{m}{1+2E}\left(  \left.\frac{r'}{r^2}\right|_- - \left.\frac{r'}{r^2}\right|_+ \right) dR + \int_0^{R_1} \frac{E'}{1+2E} \left(x^+ + x^-\right) dR  \right].
\end{equation}
Using the following identity along in- and outgoing radial null rays:
$$
\frac{d}{dR}\frac{1}{r} = \pm\frac{r'}{\sqrt{1+2E}} {\bf k}_\pm\left[ \frac{1}{r} \right]
 = -\frac{r'}{r^2} \mp \frac{r'}{\sqrt{1+2E}}\frac{\dot{r}}{r^2},
$$
integration by parts finally leads to
\begin{eqnarray*}
\frac{\nu_{\infty}^+}{\nu_{\infty}^-} &=& f_+(x_1^+) f_-(x_1^-)
 \exp\left\{ \left. \int_0^{R_1} \frac{m'}{1+2E} \left( \left.\frac{1}{r}\right|_- - \frac{1}{r}\right|_+ \right) dR   
  + \int_0^{R_1} \frac{m}{(1+2E)^{3/2}}\left( \left.\frac{r' \dot{r}}{r^2}\right|_+ + \left.\frac{r' \dot{r}}{r^2}\right|_- \right)dR \right.\\
 && \left.  + \int_0^{R_1} \frac{E'}{1+2E} \left[x^+ + x^- + \left( x^+ \right)^2 - \left( x^- \right)^2 \right] dR  \right\},
\end{eqnarray*}
with the same functions $f_\pm$ as defined in Eq.~(\ref{Eq:fpm}). The same arguments as in the marginally bound case imply that the first two integrals on the right-hand side are negative or zero. As for the third integral, it is also negative or zero according to our assumption $E' \leq 0$ and the fact that $x^+\geq x^-$. Therefore, we arrive at exactly the same estimate for the frequency ratio $\nu_\infty^+/\nu_\infty^-$ as in the marginally bound case, see Eq.~(\ref{Eq:bound}). In particular, there is a total \emph{redshift}, and for small $x_1^-$ the redshift factor scales as $(x_1^-)^3$.

\section{Numerical examples}
\label{Sec:Num}

In this section, we give some numerical examples for the redshift of a photon with zero angular momentum traveling from $\Scri^-$ to $\Scri^+$. For this, we consider the following family of initial data, corresponding to a dust cloud which is initially at rest and whose density profile is given by~\cite{nO12}
\begin{equation}
\rho_0(R) = \rho_c\left[ 1 - \left(\frac{R}{R_1}\right)^{2n} \right],\qquad
0 \leq R \leq R_1,
\label{Eq:rho0}
\end{equation}
where $R_1$ is the initial areal radius of the cloud, $\rho_c > 0$ is the central density, and $n\geq 1$ measures the flatness of the profile, the limit $n\to \infty$ corresponding to the homogeneous case discussed in Sec.~\ref{Sec:Hom}. Note that $\rho_0(R_1) = 0$ so that the density is a continuous, monotonously decreasing function on $(0,R_1)$. In the following, we focus on configurations with fixed initial radius $R_1$ and fixed total mass given by
\begin{equation}\label{Eq:Total_mass}
m_1 = \frac{4\pi}{3} R_1^3\rho_c\frac{2n}{2n + 3}.
\end{equation}
\begin{figure}[h!]
\begin{center}
\includegraphics[width=10.0cm]{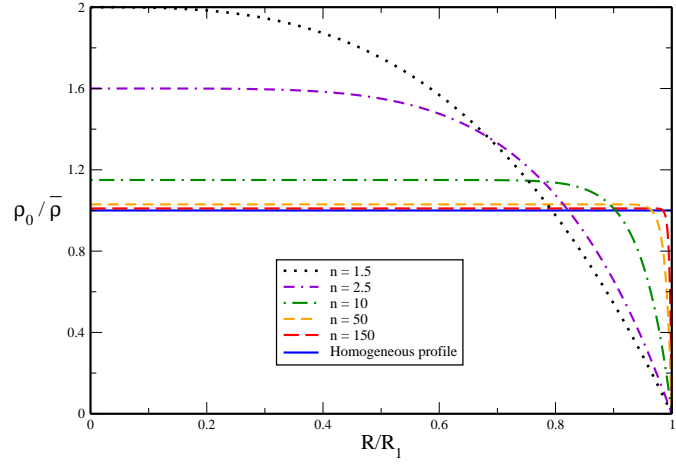}
\end{center}
\caption{The initial density profile $\rho_0(R)/\overline{\rho}$ described in Eqs.~(\ref{Eq:rho0},\ref{Eq:rhoc}) as a function of $R/R_1$ for $n=1.5, 2.5, 10, 50, 150$.
\label{Fig:c}}
\end{figure}
Accordingly, we restrict ourselves to initial density profiles of the form~(\ref{Eq:rho0}) for different values of $n\geq 1$, where the central density is given by
\begin{equation}
\rho_c = \overline{\rho}\left( 1 + \frac{3}{2n} \right),
\label{Eq:rhoc}
\end{equation}
with $\overline{\rho} := 3m_1/(4\pi R_1^3)$ the mean initial density of the cloud. The initial density profile for different values of $n$ is shown in Fig.~\ref{Fig:c}.

\begin{figure}[h!]
\begin{center}
\includegraphics[width=10.0cm]{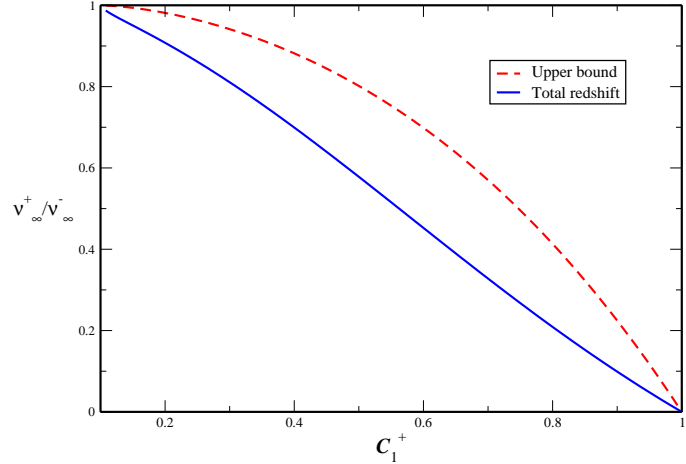}
\end{center}
\caption{\label{Fig:bound} Comparison of the total frequency shift $\nu_\infty^+/\nu_\infty^-$ with its upper bound given by Eq.~(\ref{Eq:bound}) in Sec.~\ref{Sec:TB}. The initial density profile corresponds to the one described in Eq.~(\ref{Eq:rho0}) with $n=50$ and $2m_1/R_1 =  1/10$. The bound remains valid even though in this case $E' > 0$ in a region inside the cloud close to the surface. Notice that the upper bound is exact in the limit ${\cal C}_1^+\to 1$, corresponding to a light ray exiting the cloud very close to the event horizon.}
\end{figure}
As a numerical test, in Fig.~\ref{Fig:bound}, corresponding to the profile given in Eq.~(\ref{Eq:rho0}) with $n=50$ and $2m_1/R_1 = 1/10$, we show the total frequency shift $\nu_\infty^+/\nu_\infty^-$ as a function of the compactness ratio ${\cal C}_1^+$ at which the light ray exits the cloud, and we compare this shift to the upper bound obtained in Eq.~(\ref{Eq:bound}) in Sec.~\ref{Sec:TB}. For this particular initial density profile, the condition $E'\leq 0$ fails to be valid everywhere inside the cloud. Nevertheless, Fig.~\ref{Fig:bound} shows that the bound remains valid even though the condition $E'\leq 0$ does not. The numerical integration of radial light rays was performed using a fourth-order Runge-Kutta algorithm. In the case of homogeneous density, where exact expressions for the frequency shift are known (see Sec.~\ref{Sec:Hom}), we have checked that our numerical solutions converge to the exact ones to fourth order accuracy. In the inhomogeneous case our numerical solutions were checked to be auto-convergent to fourth order accuracy also.

\begin{figure}[h!]
\begin{center}
\includegraphics[width=12.0cm]{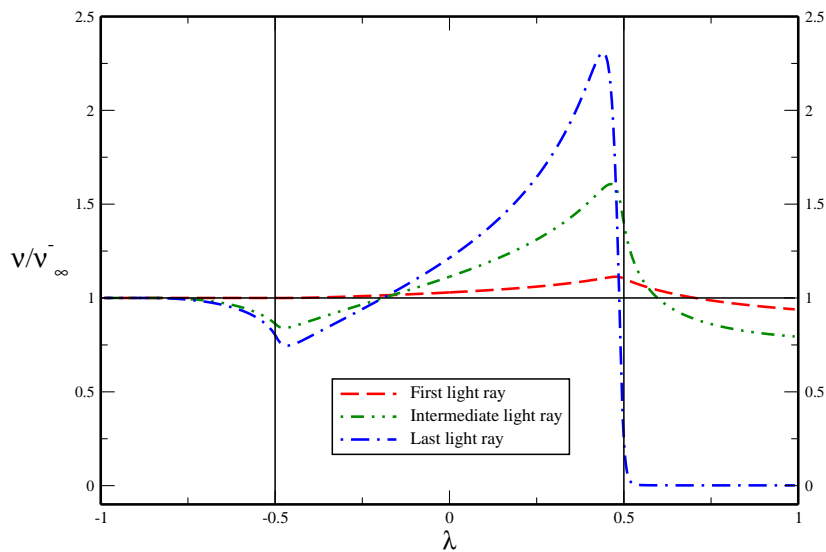}
\end{center}
\caption{\label{Fig:nu} The frequency ratio $\nu/\nu_\infty^-$ as a function of an arbitrary parameter $\lambda$ which runs from $-1$ at $\Scri^-$ to $+1$ at $\Scri^+$. The cloud corresponds to the parameter range $-0.5\leq \lambda \leq +0.5$. The results for three different light rays are displayed, the first one is the light ray penetrating the cloud at the moment of time-symmetry while the last one exits the cloud very close to the event horizon. The initial density profile corresponds to the one described in Eq.~(\ref{Eq:rho0}) with $n=10$ and $2m_1/R_1 =  1/10$. Note that the redshift in the region $-1 \leq \lambda \leq -0.5$ is due to the fact that we display the ratio $\nu/\nu^-_\infty$ measured by free falling observers instead of static ones.
}
\end{figure}

In Fig.~\ref{Fig:nu} we show the change of the frequency ratio $\nu/\nu_\infty^-$ along a radial light ray when photons travel from past to future null infinity. The frequency $\nu$ is measured by free falling observers (inside and outside the cloud). This plot shows how the photons acquire a blueshift inside the cloud, but then this blueshift is overcompensated by a redshift outside the cloud.

\begin{figure}[h!]
\begin{center}
\includegraphics[width=12.0cm]{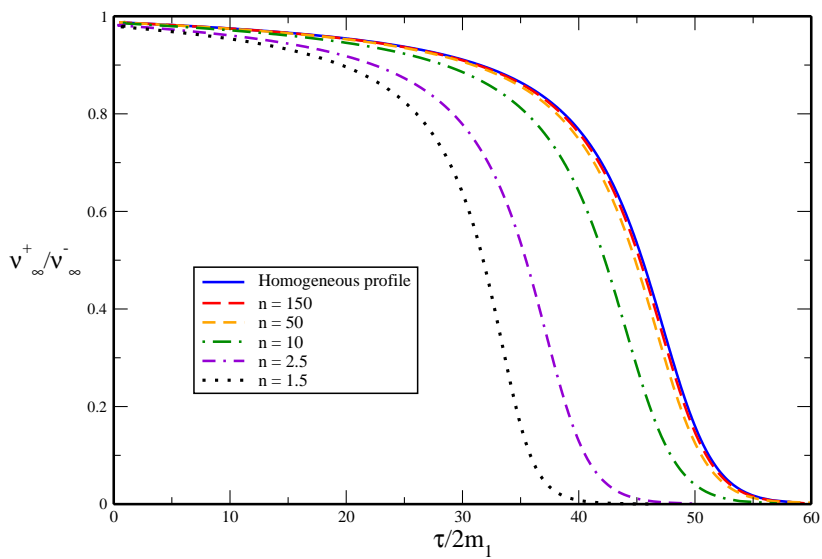}
\end{center}
\caption{The total frequency shift $\nu_\infty^+/\nu_\infty^-$ experienced by a photon traveling from $\Scri^-$ to $\Scri^+$ through the center of a cloud with fixed initial compactness ratio $2m_1/R_1 = 1/10$, as a function of proper time $\tau$ measured by a distant, static observer. This frequency shift is shown for the different density distributions displayed in Fig.~\ref{Fig:c}. We observe that for fixed mass, the redshift effect is larger for clouds whose density profile is concentrated near the center than for clouds with a nearly flat density profile.
\label{Fig:redshift}}
\end{figure}

In Fig.~\ref{Fig:redshift} we show the numerically computed frequency shift $\nu_\infty^+/\nu_\infty^-$ experienced by a photon along a light ray $\gamma$ extending from $\Scri^-$ to $\Scri^+$ through the center of a cloud for different density distributions and fixed initial compactness ratio $2m_1/R_1 = 1/10$. This frequency shift is shown as a function of proper time $\tau$ of an arbitrarily distant static observer which starts his chronometer ($\tau = 0$) when he encounters the radial light ray $\gamma_0$ that entered the cloud at the moment of time symmetry. This proper time $\tau$ can be related to the areal radius $r_1^+$ at which the photon exits the cloud in the following way. In terms of standard Schwarzschild coordinates we have
$$
\tau = \sqrt{1 - \frac{2m_1}{r_{obs}}}(t - t_0) 
 = \sqrt{1 - \frac{2m_1}{r_{obs}}}(u - u_0),
$$
where $r_{obs}$ is the radius of the static observer, $t$ and $t_0$ are the Schwarzschild times at the moment the observer crosses the light rays $\gamma$ and $\gamma_0$, and $u = t - r - 2M\log(r/2M-1)$ is retarded time which is constant along the outgoing parts of $\gamma$ and $\gamma_0$. On the other hand, at the surface of the cloud, $u$ is given by (see, for instance, Appendix C in Ref.~\cite{nOoS11})
\begin{displaymath}
u(y) = -4m_1\log(-U(y))
\end{displaymath}
where in the time-symmetric case
\begin{displaymath}
U(y) = \left( \sqrt{1 - y^2} - \frac{y}{a_1}\sqrt{1 - a_1^2} \right) 
\exp \left\{ \frac{y^2}{2a_1^2} - \frac{\sqrt{1 - a_1^2}}{2a_1^2}
\left[ \frac{1 + 2a_1^2}{a_1} \arctan \left( \frac{\sqrt{1 - y^2}}{y} \right)
 +  \frac{y}{a_1}\sqrt{1 - y^2}    \right] \right\},
\end{displaymath}
with $a_1 := \sqrt{2m_1/R_1}$ and $y := \sqrt{r/R_1}$. Since $u = const.$ along outgoing radial null rays, we obtain, for observers located at many Schwarzschild radii of the collapsing cloud ($r_{obs} \gg 2m_1$),
\begin{equation}
\tau = 4m_1\log\left( \frac{U(y_0)}{U(y_1^+)} \right),
\end{equation}
where $y_1^+ := \sqrt{r_1^+/R_1}$ and $y_0$ is the value of $y_1^+$ for the light ray $\gamma_0$. Note that $\tau = 0$ for the light ray $\gamma_0$, while $\tau\to \infty$ when $r_1^+\to 2m_1$, corresponding to the case when $\gamma$ exists the cloud very close to the event horizon.

The results in Fig.~\ref{Fig:redshift} lead to the following interesting observation. For fixed total mass $m_1$ and radius $R_1$ the redshift effect is larger for clouds whose density profile is concentrated near the center than for clouds with a nearly flat density profile. Therefore, the redshift is sensitive to the inner mass distribution of the cloud and measuring the frequency shift as a function of time could reveal information about the inner constitution of the cloud. Notice that even though the observer has to wait an infinite time ($\tau\to\infty$) to approach the horizon, the redshift effect already becomes manifest after time scales of the order of
$$
30\times 2m_1 \approx 3\times10^{-4}\left( \frac{M}{M_\odot} \right) s
$$
for the example displayed in the figure, with $M_\odot$ the solar mass.

\begin{figure}[h!]
\begin{center}
\includegraphics[width=12.0cm]{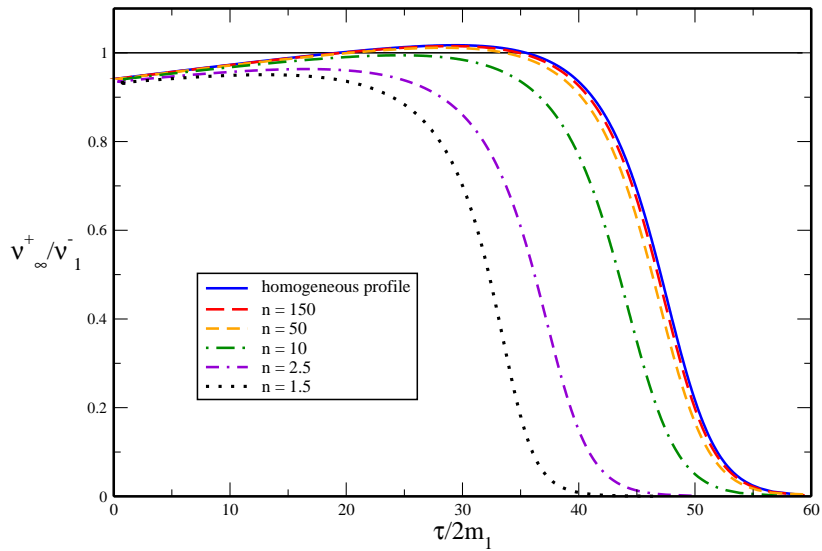}
\end{center}
\caption{Numerical computation of the frequency shift $\nu_\infty^+/\nu_1^-$ experienced by photons traveling from the surface of a cloud through its center to $\Scri^+$, as a function of proper time $\tau$ measured by a distant, static observer. In all cases the initial compactness ratio is equal to $2m_1/R_1 = 1/10$. For almost flat density profiles, some of the photons experience a blueshift. However, for non-flat profiles, there is no blueshift.
\label{Fig:blueshift}}
\end{figure}

Finally, we discuss the blueshift effect studied in Ref.~\cite{lKdMcB13}, where it was argued that photons emitted from dust particles inside the collapsing dust cloud could be \emph{blueshifted} for distant static observers. In order to discuss this point, in Fig.~\ref{Fig:blueshift}, we show the frequency shift $\nu_\infty^+/\nu_1^-$ for photons traveling on a radial light ray emanating from the surface of the cloud and traveling through the center of the cloud. As we see, for our family of initial data, there exist indeed photons which are blueshifted. However, we also see that this blueshift effect only occurs for nearly flat density profiles. For non-flat profiles $(n < 10)$ none of the photons are blueshifted, suggesting that this effect might be difficult to detect in practice.

\section{Conclusions}
\label{Sec:Conclusions}

We have studied the frequency shift of light rays traveling on an asymptotically flat dynamical spacetime describing the complete gravitational collapse of a spherically symmetric dust cloud. As a first approximation to the problem, we have assumed that the photons have zero angular momentum and do not disperse or experience any other effect due to interactions with the matter. Specifically, we have considered photons propagating from past to future null infinity through the center of the collapsing cloud and discussed the information that can be obtained from their frequency shift relative to stationary observers. The assumption that the photons do not interact with the dust particles essentially changes the conception of~\cite{lKdMcB13,lKdMcB13b}, where the photons are supposed to be emitted from the collapsing matter, involving complex processes which have to be simplified by assuming a particular spectrum of emission. In this sense, our approach is simpler and clearer, since we assume that the photons come from known sources such as supernovae explosions at small cosmological redshifts, for example.

First, we have analyzed the simple situation in which the cloud possesses an homogeneous density distribution. In this model, the collapsing cloud is described by a contracting FRW spacetime while the spacetime in the exterior is given by the Schwarzschild metric. In this scenario, we have considered photons propagating from past to future null infinity through the center of the collapsing cloud. Their frequency shift has three contributions: the first one consists of a gravitational blueshift between the static emitter near $\Scri^-$ and the surface of the cloud, combined with a Doppler redshift due to the motion of the surface relative to static observers. The second contribution is a blueshift due to the contraction of the FRW spacetime. The third contribution is a combination of a Doppler redshift at the surface of the cloud with the gravitational redshift between the surface and a static observer near $\Scri^+$. When combined together, the explicit computation of these contributions reveals that photons propagating from $\Scri^-$ to $\Scri^+$ through the center of the collapsing cloud are always redshifted with respect to stationary observers. Furthermore, we have written the frequency shift as a function of the compactness ratio of the cloud at the moment it is penetrated by the light ray. Additionally, we have explicitly calculated the frequency shift of photons which are emitted from the surface of the cloud, propagate inwards through its center and escape to $\Scri^+$. Our results confirm that some of these photons can be blueshifted as shown in Ref.~\cite{lKdMcB13} for the particular case of marginally bound collapse. However, we demonstrate that in the bounded case, there is a certain threshold for the total energy below which all the photons are redshifted.

Next, we took a step beyond the homogeneous density cloud and considered the Tolman-Bondi collapse model for generic initial data. Based on a geometric identity which relates the redshift factor along in- and outgoing radial null geodesics with the compactness ratio of the collapsing shells, we found an upper bound on the frequency shift factor of photons with zero angular momentum traveling from $\Scri^-$ to $\Scri^+$. Assuming only a nonnegative mass density, the absence of shell-crossing singularities, and the condition that the total energy of the shells is a nonincreasing function of the areal radius, this bound can be expressed in terms of the compactness ratios of the light ray at the moments it penetrates and exits the cloud. This bound implies, in particular, that these photons are always redshifted. In this context, it is also interesting to notice that the photons always experience a redshift on their trip from the center of the cloud to a distant stationary observer, as can be easily inferred from the arguments in Sec.~\ref{Sec:TB}. Therefore, the blueshift inside the cloud is dominated by the redshift in the exterior region. Although we have considered only photons with zero angular momentum in this work, we conjecture that our result is also valid for other photons traversing the dust cloud, since in this case the blueshift in the interior should be smaller. In fact, a preliminary numerical study indicates that the total frequency shift towards the red seems to prevail for photons with nonvanishing angular momentum.

We have also computed the frequency shift experienced by the photons as a function of proper time of an arbitrarily distant stationary observer and exhibited its dependency on the mass distribution of the collapsing object. This leads to an inverse problem, consisting of the determination of the mass distribution from the observed frequency shift as a function of time. The solution of this problem and its implications should be interesting.

We end this article by mentioning three further applications of the present work. The first one is related to the validity of the weak cosmic censorship conjecture. As indicated previously, the spherical dust collapse may give rise to a naked singularity even for reasonable initial data. In this case, a Cauchy horizon appears before the event horizon, extending from the first singular point of the naked singularity to $\Scri^+$. As a consequence, future null infinity of the resulting spacetime is incomplete. Then, a relevant question is whether or not the naked singularity and its associated Cauchy horizon are stable under perturbations. A first test for the stability of the Cauchy horizon is to check whether or not photons traveling arbitrarily close to the Cauchy horizon undergo an infinite blueshift, which would indicate an instability. In fact, the bound on the frequency shift derived in this article is also valid for a collapsing spherical dust cloud forming a naked singularity, as long as the compactness ratio $2m/r$ converges to zero when the center is approached along in- and outgoing light rays. Under reasonable, generic assumptions on the initial data characterizing the dust collapse this condition holds for radial light rays which pass arbitrarily close to the first singular point (see the discussion in section II of Ref.~\cite{nOoS14}). As a consequence, our bound excludes the possibility that photons traveling on such light rays are blueshifted by an arbitrary large amount, an effect that would have indicated an instability of the Cauchy horizon. This result strengthens and clarifies our earlier result in~\cite{nOoS14}, where boundedness of the blueshift in the vicinity of the singularity was shown (see also~\cite{iD98} for the marginally bound case).

The second application is again related to weak cosmic censorship and considers the possibility of testing it observationally. The idea is to compare the behavior of the redshift function found in this article for the black hole case to the case of a naked singularity. This comparison will be discussed in detail in future work~\cite{nOoStZ14}.

Finally, the total redshift effect we have discussed here could also have interesting implications in cosmology. For instance, we might consider photons emitted by distant supernova explosions or photons from the cosmic microwave background which cross several collapsing dust regions before being detected on earth. A relevant question is whether or not such photons could be redshifted by a significant amount compared to their redshift due to the expansion of the universe. In fact, there has been much recent work (see for example Refs.~\cite{nBnTeT07,jGtH08,sS11,eDmVrD12,sLcD97}) addressing related questions in the context of ``swiss cheese'' type cosmologies, where the effects of inhomogeneities on light propagation in the universe have been analyzed. It would be interesting to analyze the case in which all the inhomogeneous regions are described by massive collapsing dust clouds surrounded by an underdense region. Such regions could be modeled by a spherical dust cloud collapsing in a Schwarzschild exterior, which is precisely the situation considered in the present article except that instead of propagating from past to future null infinity, the photons would be emitted and detected within the FRW spacetime. Provided each inhomogeneous region is small compared to the Hubble scale but large compared to the extension of the collapsing cloud, such that the effects from the expansion can be neglected inside the region, the photons undergo a redshift as described in this work. Assuming that the clouds have nonrelativistic compactness ratios (${\cal C} = 2m/r \ll 1$), the calculations  in sections~\ref{Sec:Hom}~and~\ref{Sec:TB} reveal that the redshift factor $z$ due to our effect scales as
\begin{equation}
z = \frac{\nu_\infty^-}{\nu_\infty^+} - 1 \sim {\cal C}^{3/2}.
\end{equation}
It should be interesting to compare this factor to the cosmological redshift factor due to the expansion inside the region. This will be explored in future work.


\acknowledgments
We thank Luis Lehner and Thomas Zannias for suggestions and stimulating discussions. This work was supported in part by CONACyT Grants No. 46521, 232390, and 101353, and by a CIC Grant to Universidad Michoacana. Research at Perimeter Institute is supported through Industry Canada and by the Province of Ontario through the Ministry of Research and Innovation.

\bibliographystyle{unsrt}
\bibliography{../References/refs_collapse}

\end{document}